\documentclass{article}

 \usepackage[preprint]{neurips_2020}
\usepackage[utf8]{inputenc} % allow utf-8 input
\usepackage[T1]{fontenc}    % use 8-bit T1 fonts
\usepackage{hyperref}       % hyperlinks
\usepackage{url}            % simple URL typesetting
\usepackage{booktabs}       % professional-quality tables
\usepackage{amsfonts}       % blackboard math symbols
\usepackage{nicefrac}       % compact symbols for 1/2, etc.
\usepackage{microtype}      % microtypography
\usepackage{graphicx} 
 \pdfoutput=1

\title{Automated Metadata Harmonization Using Entity Resolution \& Contextual Embedding}

\author{%
  Kunal Sawarkar* \textsuperscript{1} \\
  IBM\\
  \texttt{kunal@ibm.com} \\
  \And
  Meenakshi Kodati \textsuperscript{2} \\
IBM\\
\texttt{meenakashi@ibm.com} \\
}

\begin{document}

\maketitle
\begin{abstract}
ML Data Curation process typically consist of heterogeneous \& federated source systems with varied schema structures; requiring curation process to standardize metadata from different schemas to an inter-operable schema. This manual process of \textbf{Metadata Harmonization} \& cataloging slows efficiency of ML-Ops lifecycle. We demonstrate automation of this step with the help of entity resolution methods \& also by using \textbf{Cogntive Database's Db2Vec embedding} approach to capture hidden inter-column \& intra-column relationships which detect similarity of metadata and then predict metadata columns from source schemas to any standardized schemas. Apart from matching schemas, we demonstrate that it can also infer the correct ontological structure of the target data model. 
\end{abstract}

\section{Introduction}

Large multinational organizations that gather data from disparate data sources are often faced with the challenge of standardizing the data formats to make the datasets inter-operable. This involves finding similarities among data collection methodologies and creating mapping tables that meaningfully unite information from them. This method of mapping identical or equivalent metadata across varied datasets is known as \emph{Metadata Crosswalking or Master Metadata Synchronization or Metadata Harmonization}. This includes task of metadata mapping \& cleansing the metadata to given standard and linking it to a standard structure. For a large dataset which has hundreds of columns, mapping metadata to a catalog can be a big challenge. 

One of the limitations of the data curation \& cataloging tools available today is that they are not equipped with metadata harmonization capabilities that automate the process of crosswalking. In such cases organisations typically rely on Data Stewards to perform this task manually, often consuming several days or weeks of concentrated effort. Data Stewards manually contrast each column and map it to a set standard schema. This being first step in curation, also becomes bottlneck for efficincy of data science project executions. There is also a possibility that the new data proposes additions or changes to the current ontological structure that the standard schema has to follow. \\

%First Set of Diagram

\begin{figure}[h]
	\begin{minipage}{.6\textwidth}
		\includegraphics [width=0.75\linewidth,height=0.15\textheight] {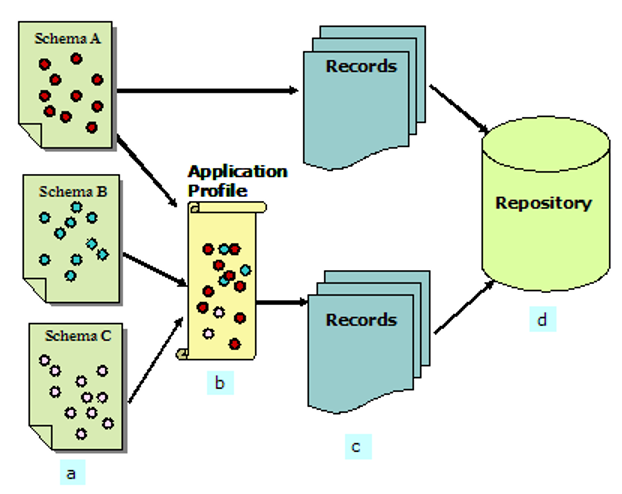}
		\caption{Crosswallking metadata for multiple schemas}
		%label{Crosswallking metadata for multiple schemas}
	\end{minipage}
	\begin{minipage}{.6\textwidth}
		\includegraphics [width=0.55\linewidth,height=0.15\textheight] {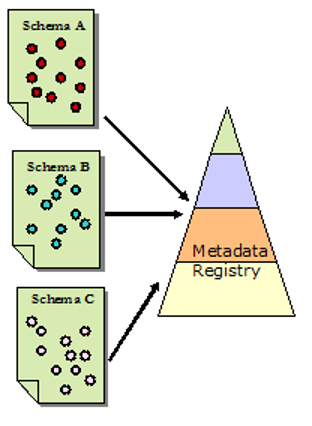}
		\caption{Mapping ontological structure to standard metadata}
		%\label{Mapping ontological structure to standard metadata}
	\end{minipage}
\end{figure}

We tacked above problems by asking ; Can we have a way we can augment data curation process with the ability to automatically crosswalk the metadata to a standard format at the time of ingestion? Also how can we derive an ontological metadata structure that can be scaled to a wide variety of data sources to automatically enhance the metadata of the datasets? And how to achieve both together?

We show that our method of metadata harmonization \emph{(1) Can automatically crosswalk the metadata from different sources which has collected the data from different methodologies and classification to a given standard schema. (2) It can also autonomously derive the standard ontological structure of metadata from the multitude of the source systems. Our method makes use of Machine learning based Entity Resolution Methods using Levenstein distance-based mechanism to find the relevent entities using cost-based approaches. (3) In further experimentation we found that by using embedding methods like db2vec of contextual vectorization and textification we resolve metadata entities to the nearest standardized schemas.} The system works well for inferring metadata column names in a schema as well ontological hierarchical structure for that schema in experimentation performed on test data. 

\section {Related Work}
 We could not find other similar application \& consider this to be an innovative use of ML to automate \& improve ML-ops process of cataloguing. In previous attempts by Joerg Wurzer[1] system harmonizes dataset with superordinate abstract data model. However key limitation of this work is that it uses predefined semantic middleware which works like static data dictionary to map metadata. Other effort in this area by Andrew Schon[2] included creating business rules for the mapping; which again are static \& not adapatable to changing needs, while work by Bill Cope[3] combines business rules \& data dictionary into interlanguage document type definition (DTD) to identify the crosswalk for metadata schemas. But all such systems require apriori knowledge as well as manual effort and does not learn to predict for newer schemas or various ontologies of schemaset automatically. 

\section{System Design}

\subsection{Framework Overview}

\begin{figure}[h]
	\includegraphics[width=0.9\linewidth, height=0.2\textheight]{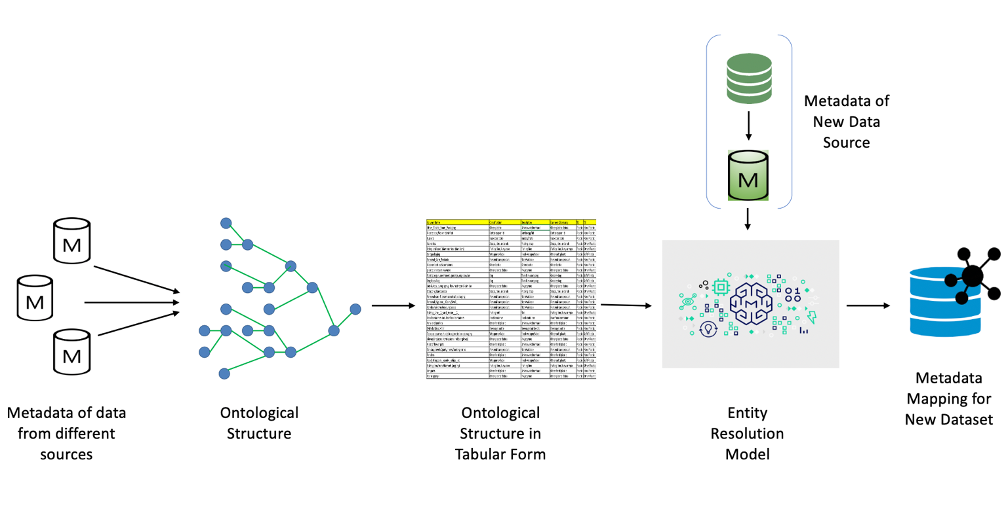}
	\caption{Framework for the Metadata Harmonization using Entity Resolution Mechanism}
\end{figure}

The framework can be summarised as below. 

1. The metadata for the crosswalk from various heterogenous sources is collected and arranged in an ontological structure. It also includes additional meta-metadata like lookup for a column attributes for a given standardization format and a data model schema structure with its own naming \& descriptive structure for which crosswalk is expected to be performed.  All the above metadata is stored in a data frame which can include source column names and the expected standardized schema tiers or data model tiers for the crosswalk output. 

2. Based on this meta-metadata like column name and the descriptions the metadata entity are resolved using machine learning methods for the crosswalk predictions. Various embodiments can be used for entity resolution to perform schema crosswalk

\textbf{a. Levenstein Distance Based Method}- This method uses the cost based textual distance to find entity matching on each column names and to the standardized schema of crosswalk model. For tie resolution in case of same Levenstein scores a blocked indexing method is used including hybrid method. 

\textbf{b.Using Contextual Embeddings}- Using \emph{db22vec} which captures sematic context for the entities and find similar entities to the crosswalk. Further we found that using method  \emph{db2vec} [4], it creates textification and then vectorization for all the entities match the entities to the expected schemas by capturing column relationship.

3. In case there exists a ground truth for the entity matching from previous manual efforts done by data stewards then an additional text classification model can learn from previous metadata crosswalks and improve the performance. The system can also work in the absence any training data to learn explicitly context of column relationships.The solution methodology can also derive \& learn metadata from cannonical industry data models to specific domain (like healthcare, banking etc.). \\

\subsection{Implementation}

The first step in the process is to obtain the ontology that can be used for the Entity Resolution model. Each metadata entity is described by the meta-metadata which can includes any of the below descriptive information about columns viz. \emph{a.	Verbose Column Name, b.	Data classification Business Terms, c.	Any Textual Description of the column, d.	Business Glossary, e.	Data Dictionary}. We can use one of the above descriptive characteristics or any combination of therein. In cases where a large number of datasets are available to begin with, the datasets chosen to create the ontology need to come from diverse data sources and should encompass a wide array of possible column names in order be effectively used or crosswalking new datasets. The standard schema can further be refined to standardize column name formats, remove erroneous data and eliminate duplicate values.

\subsubsection{Approach1: Using Levenshtein Distance for Entity Resolution Model}

In order to perform entity resolution on newly ingested data, we first tried Levenshtein distance as a metric for measuring the distance between two metadata strings. The score ranges between 0 and 100, with 100 being an indication for exact match between the column strings. In addition to deciding on the method for derivation of the matching score, the minimum matching score that indicates a ‘qualified match’ was chosen to be above 70. In other words, the threshold score, crossing which a pair of strings can be considered to be ‘closely matching’ needs to be set. This naive approach worked well for column metdata match but did not always capture context of inter-column relationships \& ontology in complex cases.

\begin{figure}[h]
	\begin{minipage}{.6\textwidth}
		\includegraphics [width=0.8\linewidth,height=0.2\textheight] {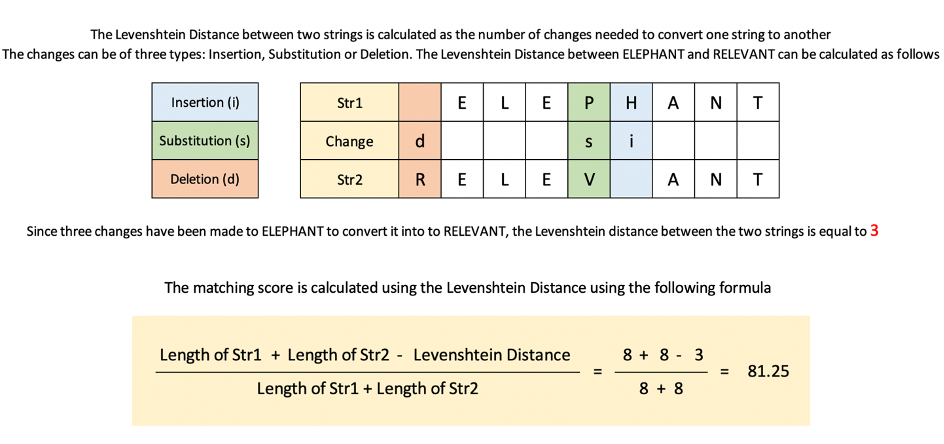}
		\caption{Using Levenshtein Distance }
		%label{Crosswallking metadata for multiple schemas}
	\end{minipage}
	\begin{minipage}{.6\textwidth}
		\includegraphics [width=0.99\linewidth,height=0.2\textheight] {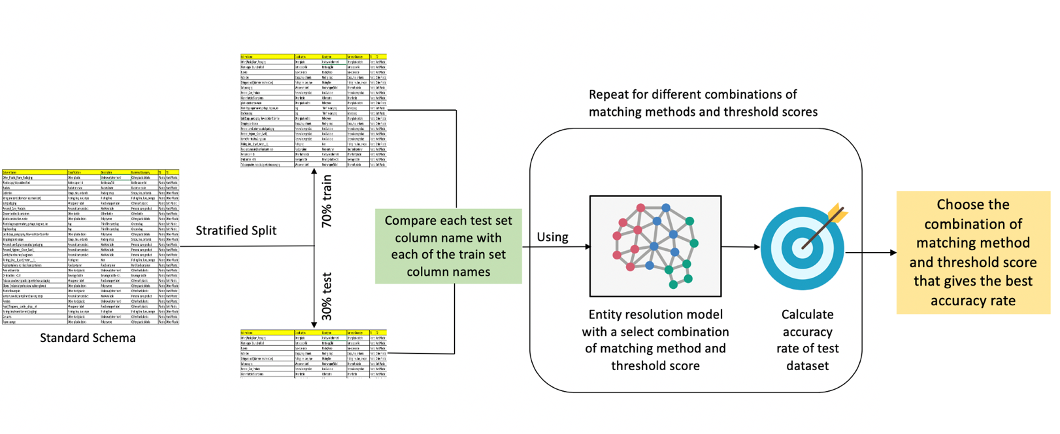}
		\caption{Using db2vec method for Embeddings}
		%\label{Mapping ontological structure to standard metadata}
	\end{minipage}
\end{figure}

\subsubsection{Approach 2: Using db2vec method for  Embeddings}

Thus we applied Db2Vec embedding from cognitve database systems [4] that captures the inter column \& intra column relationship between tables by treating it as unstructured set. The first step in this process is the textification of the Standard Schema ontology. When the standard schema is provided as an input in the form of a dataframe, the algorithm first textifies the data (using a preprocessing script to convert every row of the table into the required format sentences).The resulting textified standard schema is then used to train the db2vec model. The model takes the textified data as input and outputs a vector representation of the unique tokens in the textified data. The training involves the use of a highly specialized 3-layer Neural Network Model to generate vectors. Once trained, this model can take an unknown string as an input and provide a list of records from the standard schema that closely match the input string. In this use case, the input string is a column name from a newly ingested dataset. This model is adept at finding similar records even when there are errors or spelling mistakes in the input column. The train set can now be replaced with the complete standard schema, and metadata of a new, unseen dataset can now take place of the test set. The model would compare each column metadata of the new dataset against the standard schema to find the closest match.  The ontological structure assigned to the best matching column name would then be picked from the standard schema and assigned to the new column, thereby, rendering the model the capability to crosswalk.

\section{Experimentation}

We performed experiment on the open data systems for Marine Litter managed by United Nations Env Program (UNEP) \& Earth Challenge. All these datasets have different data collection methodology, different schemas and is not interoperable[5,6,7]. A team of Data Stewards had spent lot of manual effort to harmonize metadata and created a inter-operable dataset file with uniform metadata[8]. We tested our method to these source sets and compared the results of our model to manually curated metadata.\\

 \begin{figure}[h]
	\begin{minipage}{.57\textwidth}
		\includegraphics [width=0.99\linewidth,height=0.2\textheight] {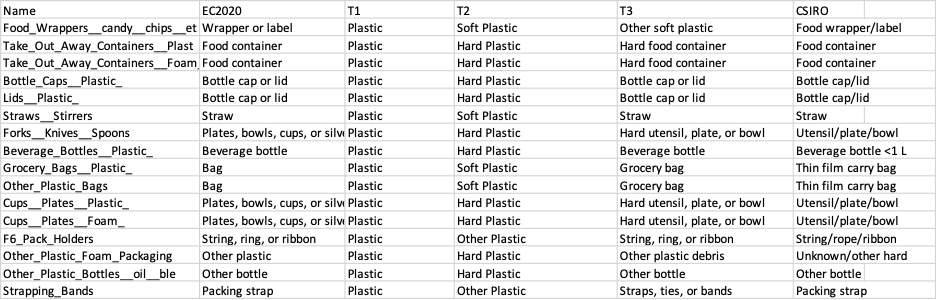}
		\caption{Example of source schemasets}
		%\label{Crosswallking metadata for multiple schemas}
	\end{minipage}
	\begin{minipage}{.7\textwidth}
		\includegraphics [width=0.9\linewidth,height=0.2\textheight] {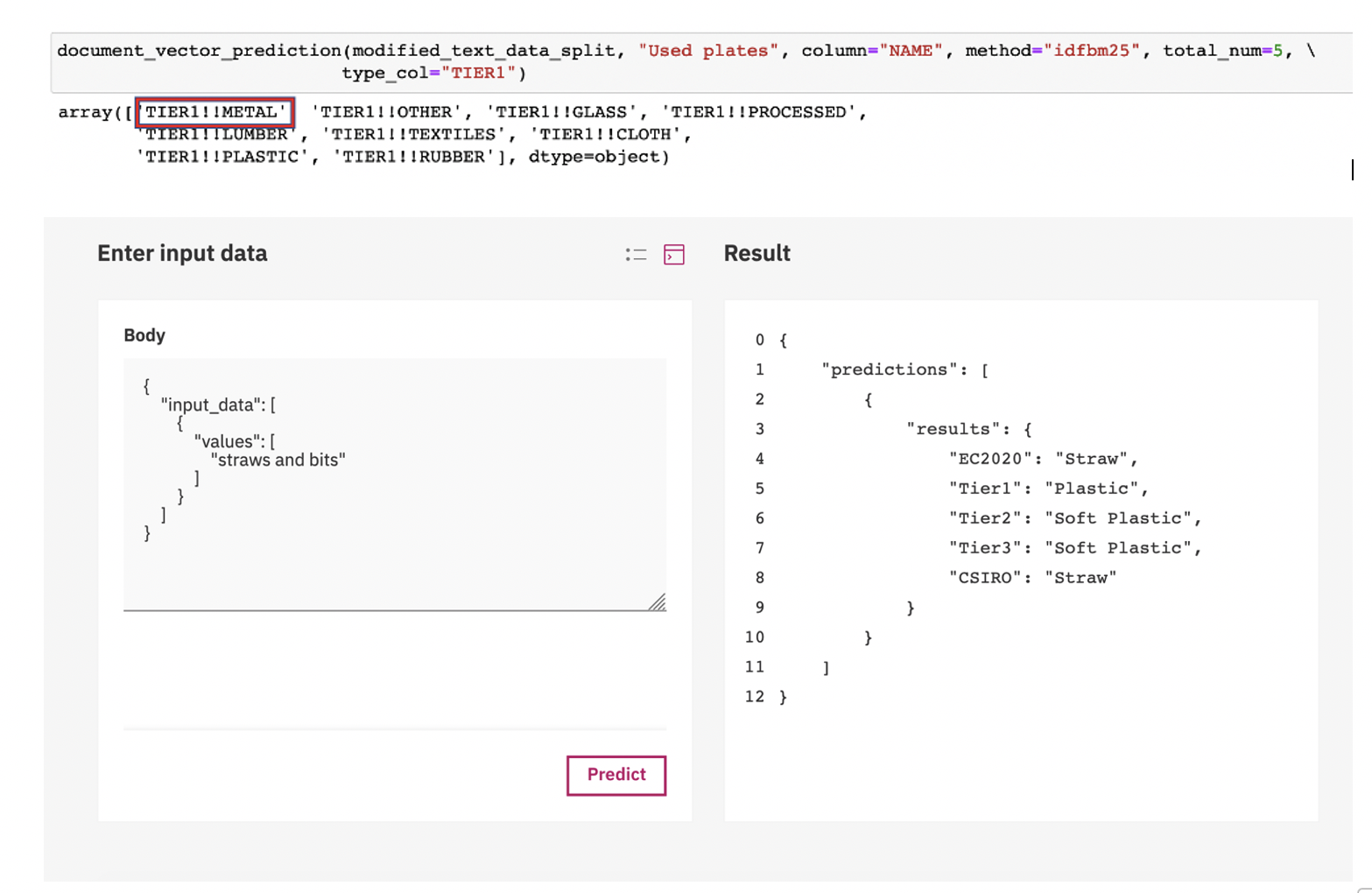}
		\caption{Predicting column metadata (above) \& ontology (below)}
		%\label{Mapping ontological structure to standard metadata}
	\end{minipage}
\end{figure}

Example- For marine litter, the classification of plastic litter itself can be done in lots of different manner by various schemas which requires manual analysis to map them with each other. T1 , T2, T3 refpresent different organization schemas for same object. In test result, when the \emph{Marine Debris }data is provided with just one level of classification \emph{Tier1}, the top result for a new input, \emph{Used Plates}, correctly came out to be \emph{Metal}. It also predicted ontological hierachy for \emph{straw} as \emph{soft plastics}. We also ran experiment on completely new schemas.We observed overall accuracy of 82 percent for this dataset consiting few hundred columns in each of the child schemas.

\section{Conclusion}
We present a novel approach to harmonize metadata automatically by creating a machine generated crosswalk. We further demonstrated that, it can work in the absence any training data using contextual embeddings of entities and also that the system can derive not just column names but also the ontological structure of schema. We thus show that our system can speed-up data curation pipeline.

\pagebreak

\section* {Statement on Impact}

This research is primarily aimed at influencing the efficiency of ML ops process in large enterprises. As popular wisdom  goes; 80 percent of Machine Learning project effort is spent just on data engineering and within  that majority of effort is on cataloguing dataset for data curation by harmonization various schemas. Application of ML for improving ML lifecycle itself is the intent of the authors and we believe this shall greatly benefit data stewards who are mainly responsible for this work. This will also benefit data engineers  data scientist by accelerating the data curation process. Authors currently do not foresee disadvantages of this method to any particular group nor see the implications of bias. Authors do understand that this research may change the nature of job for the community of Data Stewards but this should be viewed as advantageous to them by improving productivity of tasks instead of being a disadvantage or potential risk to job. Authors welcome any constructive feedback on the impact of this research which they may not have anticipated. \\

\section*{References}

\medskip

\small

[1] Joerg Wurzer \ (2013) {Automated harmonization of data }

[2] Andrew Schon\ (2011) {Matching metadata sources using rules for characterizing matches}.

[3] Bill Cope\ (2003)  {Method and apparatus for the creation, location and formatting of digital content} 

[4] Bordawekar, Rajesh \& Bandyopadhyay, Bortik \& Shmueli, Oded. (2017). {Cognitive Database: A Step towards Endowing Relational Databases with Artificial Intelligence Capabilities}. 

[5] [Data] Trash Information and Data for Education and Solutions (TIDES): Plastic Pollution \emph{https://cscloud-ec2020.opendata.arcgis.com/datasets/data-marine-litter-watch-mlw-plastic-pollution-}

[6] [Data] Marine Litter Watch (MLW): Plastic Pollution \emph{https://cscloud-ec2020.opendata.arcgis.com/datasets/data-marine-litter-watch-mlw-plastic-pollution-}

[7] [Data] Marine Debris Monitoring and Assessment Project (MDMAP) Accumulation Report: Plastic Pollution - \emph{https://cscloud-ec2020.opendata.arcgis.com/datasets/data-marine-debris-monitoring-and-assessment-project-mdmap-accumulation-report-plastic-pollution}

[8] [Data] Earth Challenge Integrated Data: Plastic Pollution (MLW, MDMAP, TIDES)  \emph{https://cscloud-ec2020.opendata.arcgis.com/datasets/data-earth-challenge-integrated-data-plastic-pollution-mlw-mdmap-tides-}

\end{document}